\documentclass[namedreferences]{SolarPhysics}
\usepackage[optionalrh]{spr-sola-addons} 
\usepackage{epsfig}                     
\usepackage{graphicx}                    
\usepackage{multirow}
\usepackage{color}                       
\usepackage{url}                         
\usepackage{rotating}

\begin{document}

\begin{article}

\begin{opening}

\title{Effect of Solar Wind Drag on the Determination of the Properties of Coronal Mass Ejections from Heliospheric Images}

%
\author{N.~\surname{Lugaz}$^{1}$\sep P.~\surname{Kintner}$^{1,2}$}

%
\runningauthor{Lugaz and Kintner}
\runningtitle{Effect of Drag on Determining CME Properties}

%
  \institute{$^{1}$ Institute for Astronomy - University of Hawaii-Manoa, 2680 Woodlawn Dr., Honolulu, HI 96822, USA
                     email: \url{nlugaz@ifa.hawaii.edu}\\ 
$^{2}$ University of Rochester, 
                     email: \url{pkintner@u.rochester.edu}}

\begin{abstract}

The Fixed-$\Phi$ (F$\Phi$) and Harmonic Mean (HM) fitting methods are two methods to determine the ``average'' direction and velocity of coronal mass ejections (CMEs) from time-elongation tracks produced by Heliospheric Imagers (HIs), such as the HIs onboard the STEREO spacecraft. Both methods assume a constant velocity in their descriptions of the time-elongation profiles of CMEs, which are used to fit the observed time-elongation data. Here, we analyze the effect of aerodynamic drag on CMEs propagating through interplanetary space, and how this drag affects the result of the F$\Phi$ and HM fitting methods. A simple drag model is used to analytically construct time-elongation profiles which are then fitted with the two methods. It is found that higher angles and velocities give rise to greater error in both methods, reaching errors in the direction of propagation of up to 15$^\circ$ and 30$^\circ$ for the F$\Phi$ and HM fitting methods, respectively. This is due to the physical accelerations of the CMEs being interpreted as geometrical accelerations by the fitting methods. Because of the geometrical definition of the HM fitting method, it is affected by the acceleration more greatly than the F$\Phi$ fitting method.  Overall, we find that both techniques overestimate the initial (and final) velocity and direction for fast CMEs propagating beyond 90$^\circ$ from the Sun-spacecraft line, meaning that arrival times at 1 AU would be predicted early (by up to 12 hours). We also find that the direction and arrival time of a wide and decelerating CME can be better reproduced by the F$\Phi$ due to the cancellation of two errors: neglecting the CME width and neglecting the CME deceleration. Overall, the inaccuracies of the two fitting methods are expected to play an important role in the prediction of CME hit and arrival times as we head towards solar maximum and the STEREO spacecraft further move behind the Sun.

\end{abstract}
\end{opening}

%
%

\section{Introduction} \label{intro}

Coronal mass ejections (CMEs) are large-scale eruptions of plasma from the Sun and are one of the main drivers of space weather, accounting for about 85$\%$ of intense geomagnetic storms \cite{Zhang:2007,Echer:2008}. The occurrence rate of CMEs vary with the 11-year solar cycle, from no more than one event a day near solar minimum to as many as 6--8 CMEs per day at solar maximum \cite{Riley:2006b}. Statistical studies have shown that about 70$\%$ of front-sided halo CMEs observed by coronagraphs result in a geo-effective ejecta at 1~AU \cite{Gopalswamy:2007}, while about 64$\%$ of ejecta at 1 AU for which a source region was identified are associated with CMEs within 20$^\circ$ of central meridian \cite{Richardson:2010}. Different methods have been devised to predict whether a CME will hit Earth or not and, if yes, to forecast the arrival time. Until the launch of the {\it Solar-Terrestrial Relation Observatory} (STEREO) in 2006, most of these methods relied on information from coronagraph measurements, the flare location and, if available radio bursts. Empirical models, such as the Empirical Shock Arrival (ESA, see: \opencite{Gopalswamy:2001b}, \opencite{Gopalswamy:2005}), have been able to predict the arrival time of CMEs with typical errors of $\pm$ 12 hours. 

The twin STEREO spacecraft \cite{Kaiser:2008}, launched in November 2006, have given us continuous views of CMEs from the solar surface out to 1~AU via the SECCHI suite \cite{Howard:2008} through the use of two coronagraphs (COR-1 and COR-2) and two Heliospheric Imagers (HI-1 and HI-2 or HIs collectively). There are two copies of STEREO, A and B, both spacecraft orbit the sun at roughly 1~AU moving away from the Sun-Earth line at about 22.5$^\circ$ per year; STEREO-A orbits ahead of Earth's orbit  and STEREO-B orbits behind it. By combining the two HIs, the total field-of-view of the heliospheric imagers range from about 4$^\circ$ to 82$^\circ$ \cite{Eyles:2009}. Real-time observations from the HI instruments can be used to predict the arrival time of CMEs using highly-compressed ``beacon'' data \cite{Davis:2011,THoward:2010}. Preliminary studies show that the typical error in the predicted arrival time is reduced to about $\pm$ 6 hours. Analyses of the high-resolution scientific data have shown that the arrival times for Earth-directed CMEs can be reproduced with even lower errors using a range of techniques \cite{Davies:2009,Wood:2009b,THoward:2009a,Rouillard:2009b,Liu:2010b,Lugaz:2010b,Lugaz:2010c}.

One of these families of techniques is based on fitting methods following the work of \inlinecite{Sheeley:1999}. The central concept of these techniques is that, assuming a constant velocity and constant direction of propagation, CMEs appear as decelerating (resp. accelerating) as they propagate away (resp. towards) the observing spacecraft. This apparent deceleration or acceleration can be explained purely from geometrical considerations. Different assumptions regarding the geometrical shape of the CMEs can be made, resulting in different techniques. The Fixed-$\Phi$ (F$\Phi$) fitting method assumes that the nose of the CME is what is being imaged at all times by the HIs \cite{Rouillard:2008}. The Harmonic Mean (HM) fitting method assumes that the CME front can be locally described as a circular arc attached to the Sun \cite{Lugaz:2009c}. Then, the HIs do not necessarily observe the nose of the CME but instead the tangent to this CME front \cite{Lugaz:2010c}. Both methods fit observed time-elongation data to analytical functions expressing the elongation angle with respect to time depending on the CME direction and speed. The two methods were recently compared for some Earth-directed CMEs \cite{Moestl:2011} as well as STEREO-impacting CMEs \cite{Lugaz:2012a} and for all the tracks observed by the HIs from April 2007 to September 2011 \cite{Davies:2012}.

At the core of the HM and F$\Phi$ techniques is the assumption that an ``average'' speed and direction can be found which characterize the CME propagation. Previous work have shown that deflection of CMEs can be strong in the corona \cite{Gui:2011,Kilpua:2009b,Byrne:2010} whereas it is often negligible in the heliosphere \cite{Lugaz:2010b,Liu:2010b}. The best-fit speed has been used to predict the CME speed at 1~AU \cite{Moestl:2011,Rollett:2011,Davis:2011}; however, unless the CME speed is indeed constant, the average transit speed should not necessarily match the final speed at 1~AU. 

CME propagation is affected by the Lorentz force, gravity, and aerodynamic drag, however past 20~R$_\odot$, the effect of the Lorentz force and gravity are thought to be negligible \cite{Vrsnak:2006,Vrsnak:2010}. This implies that the dominant force experienced by CMEs is the drag force from the solar wind. The drag arises from CME interaction with the solar wind, transferring momentum to either speed up or slow down the CME velocity so that it approaches the solar wind speed \cite{Vrsnak:2010, Richardson:2010}. Previous studies with data from SECCHI and from the Solar Mass Ejection Imager (SMEI) have tried to fit CME reconstructed position from white-light measurements with kinematic equations \cite{Chen:1996,Cargill:2004,Tappin:2006} taking into account the solar wind drag and the Lorentz force \cite{THoward:2007,Webb:2009}. These studies proved that it is possible to match observations with different physical models. However, the uncertainty on the CME and model  parameters (drag coefficient, CME direction, CME mass) and reconstruction techniques is large enough that it is usually not possible to distinguish between different physical models. Recently, \inlinecite{Temmer:2011} combined {\it in situ} measurements about the arrival time and final speed of CMEs with observational data from SECCHI to determine the distance at which the CME drag becomes the dominant force acting on CMEs. 

Here, we analyze how the varying velocity of a CME affects the fitting methods commonly used to analyze HI observations. While statistical studies based on real data similar to \inlinecite{Lugaz:2010c}, \inlinecite{Davies:2012} and \inlinecite{Lugaz:2012a} would be an optimal way to tackle this problem, there has not been enough fast CMEs to perform such a study. In addition, a study based on real observations would need to account for the varying solar wind and observational conditions. Therefore, it is somewhat easier to base such a study on analytical or computational methods in order to ``control'' the background solar wind conditions and the STEREO observational viewpoint.  In a previous study \cite{Lugaz:2011b}, we performed a numerical simulation of a moderately fast CME ($\sim$ 800~km~s$^{-1}$) with the Space Weather Modeling Framework \cite{Toth:2005} and we found that the deceleration yields larger errors for larger viewing angles and can reach $\sim 5^\circ$ for viewing angles of 75$^\circ$. Here, we expand this study to cover more CME speeds and larger viewing angles. In section~2, we discuss the method used to create time-elongation profiles of CMEs and we give a quick overview of the fitting methods tested. In section~3, we first give an example of our analysis for a profile corresponding to a fast CME observed behind the limb before discussing the general results. In section~4, we draw our conclusions and discuss consequences for future observations of CMEs and the use of fitting methods for space weather forecasting.

\section{Drag Equation and Fitting Methods} \label{methods}

In order to test the effect of solar wind drag on the derivation of properties of CMEs, we first create synthetic time-elongation profiles taking into account the interaction of the CME with the solar wind by solving for the CME position and taking into account a drag term. Then, we fit these profiles with the two common fitting methods to get the CME speed and direction. Below, we outline the procedure to create and fit the time-elongation profiles.

\subsection{Fitting Methods}

When CMEs are observed to large elongation angles, the time-elongation data can be fitted to analytical functions and a single value for the CME speed and direction of propagation can be derived \cite{Sheeley:1999}. The two most common methods are the Fixed-$\Phi$ (F$\Phi$) fitting of \inlinecite{Rouillard:2008} and the Harmonic Mean (HM) fitting of \inlinecite{Lugaz:2010c}. Here, we use the following notation: $\alpha$ is the elongation angle, $\Phi$  the direction of propagation,  which is assumed to be fixed, $d_\mathrm{ST}$ the heliocentric position of the observing spacecraft (STEREO), $t$ the time and $V$ the ``average'' speed.
 
Assuming that what is observed is a single plasma element, the relation between elongation and distance of \inlinecite{Kahler:2005} can be inverted to obtain the F$\Phi$ fitting relation \cite{Rouillard:2008}:

\begin{equation}
\alpha = \arctan \left (\frac{ V  t \sin\Phi }{d_\mathrm{ST} - V t \cos\Phi} \right).\label{eq:TEFF}
\end{equation}
  
Another simple assumption for what is observed by HIs, as proposed by \inlinecite{Lugaz:2009c} and \inlinecite{THoward:2009a}, is as follows. The CME front can be modeled as a locally circular front  with a diameter equal to the distance of the apex (or equivalently, a front which is anchored at the Sun). It is further assumed that the measured elongation angle simply corresponds to the angle between the Sun-spacecraft line and  the line-of-sight tangent to this circular front. Assuming a constant propagation speed and the geometry explained above, the HM fitting relation can be obtained \cite{Lugaz:2010c,Liu:2010b,Moestl:2011}:

\begin{eqnarray}
\alpha &= & \arctan \left (\frac{ V  t \sin\Phi}{2 d_\mathrm{ST} - V t \cos\Phi} \right) +  \label{eq:TEHM} \\
& & \arcsin \left (\frac{ V t  }{\sqrt{\left(2 d_\mathrm{ST} - V t \cos\Phi \right)^2 + \left(V  t \sin\Phi \right)^2}} \right). \nonumber
\end{eqnarray}

A measured time-elongation profile can be fitted to profiles of calculated elongations  given by Equations~(1) and (2). 

\subsection{Drag Equation and Time-Elongation Data}

In order to create synthetic time-elongation profiles, the first step is to create time-distance profiles for hypothetical CMEs in presence of a drag term. The force experienced by CMEs due to their interaction with the solar wind depends on a large number of factors, including the CME mass and cross-section area, the solar wind density and speed \cite{Cargill:2002,Vrsnak:2010}. Here, we use a simplified form of the acceleration (force by unit mass) following \inlinecite{Maloney:2010} and \inlinecite{Vrsnak:2002}:
\begin{equation}
a = - C r^{-\frac{1}{2}} |V - V_\mathrm{sw} | \left ( V - V_\mathrm{sw} \right),
\end{equation}
where $a$ is the acceleration, $V_\mathrm{sw}$ is the solar wind speed, $r$ the heliocentric distance and $C$ is a constant related to the drag coefficient.

Since we only focus on the propagation of the CMEs in the HIs field-of-view, beyond 0.1~AU, we neglect the effect of the gravitational force as well as that of the Lorentz force (see \opencite{Temmer:2011} for a discussion of the distance at which the drag term becomes the dominant force acting on CMEs). We further assume that the solar wind speed is constant at this distance to a value of 320~km~s$^{-1}$, typical of slow solar wind conditions. We used for $C$ one of the values given in \inlinecite{Maloney:2010}: 1.28 $\times 10^{-7}$ (SI units). We chose this value as it gives kinematics and transit time typical of real CMEs: a CME with an initial speed (at 0.1~AU) of 800 km~s$^{-1}$ takes approximately 60 hours to propagate from 20~R$_\odot$ to 1 AU, whereas a CME with an initial speed of 1800 km~s$^{-1}$ takes about 29 hours, value in agreement with typical observations as well as  empirical models \cite{Gopalswamy:2005, Richardson:2010}. Overall, we are left with a simple differential equation which links the CME speed to the heliocentric distance. 
\begin{equation}
\frac{V\,dV}{ |V - V_\mathrm{sw} | \left ( V - V_\mathrm{sw} \right)} = - C r^{-\frac{1}{2}}  dr
\end{equation}
A similar equation can be found in \inlinecite{Vrsnak:2002} and \inlinecite{Byrne:2010}. We discuss the possible errors associated with our assumption in section~4. 

Assuming initial velocities of 200, 500, 800, 1200, and 1800~km~s$^{-1}$, Equation~(4) is solved using a fourth-order Runge-Kutta method from 20~$R_\odot$ to 1~AU. This gives profiles of velocity as a function of distance, which are then integrated for time to provide profiles of distance as a function of time. Assuming viewing angles of  15$^\circ$, 30$^\circ$, 45$^\circ$, 60$^\circ$, 75$^\circ$, 90$^\circ$, 105$^\circ$ and 120$^\circ$, we derive synthetic time-elongation profiles for each combination of velocity and viewing angle, using Equations (1) and (2) for the F$\Phi$ and HM methods respectively. Finally, a sample of points is taken from each profile at equal temporal intervals of 1.5 hours, resulting in 20--100 datapoints for each profile, depending on the CME initial velocity. For simplicity sake, no sampling error was added onto the profiles. Manual sampling has been shown to result in errors of about 2--5$^\circ$ \cite{Williams:2009}. Because the synthetic time-elongation profiles are created directly using the HM and F$\Phi$ equations, there is no error associated with the geometry. Fitting these tracks with the same equations, the difference between the viewing angle and the best-fit angle is purely due to the acceleration/deceleration of the CME as given by Equation~(4). As we did in \inlinecite{Lugaz:2010c}, we also fit the F$\Phi$ (resp. HM) with the HM fitting method (resp. F$\Phi$ method) to see how the different errors combine or cancel each other.

\section{Results}

\subsection{Example of a Fast CME Observed Behind the Limb}

Here, we give an overview of our analysis for two synthetic profiles corresponding to the case of a fast CME observed behind the limb: viewing angle of 120$^\circ$ corresponding to the separation between Earth and STEREO-A in July 2012 and between Earth and STEREO-B in October 2012. Some STEREO-directed CMEs (for example the 2009 December 4 CME) have been observed remotely beyond 0.5~AU with such large viewing angles \cite{Lugaz:2012a}. We expect that 120$^\circ$--130$^\circ$ is the largest viewing angle for which Earth-directed CMEs will be observed remotely to large enough distances to successfully apply fitting techniques. Figure~1 shows the distance-time and velocity-time profiles of a CME with initial speed at 0.1~AU of 1200~km~s$^{-1}$. Note that there were at least 15 CMEs with similar or faster speeds observed by LASCO each year from 1999 to 2004, half of which were typically halo CMEs. Therefore, we can expect that SECCHI will observe a few (2 to 10) Earth-directed CMEs this fast in 2012. With our solar wind drag model, such a CME has a transit time from 20~R$_\odot$ to 1~AU of 41.8 hours, a final speed of 759~km~s$^{-1}$ and an average transit speed of 901~km~s$^{-1}$.

\begin{figure*}[t*]
\begin{center}
{\includegraphics*[width = 8.5cm]{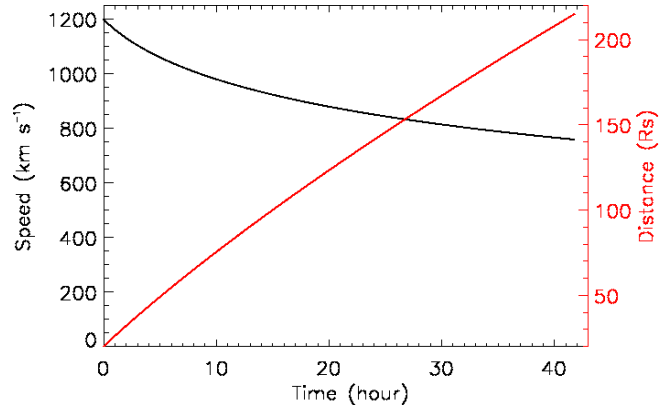}}
\caption{Time-distance (red curve, right scale) and time-velocity (black, left scale) profiles for a CME with an initial speed at 0.1~AU of 1200~km~s$^{-1}$ and with an evolution subject to the drag term described in the text.}
\end{center}
\end{figure*}

\begin{figure*}[t*]
\begin{center}
{\includegraphics*[width = 8cm]{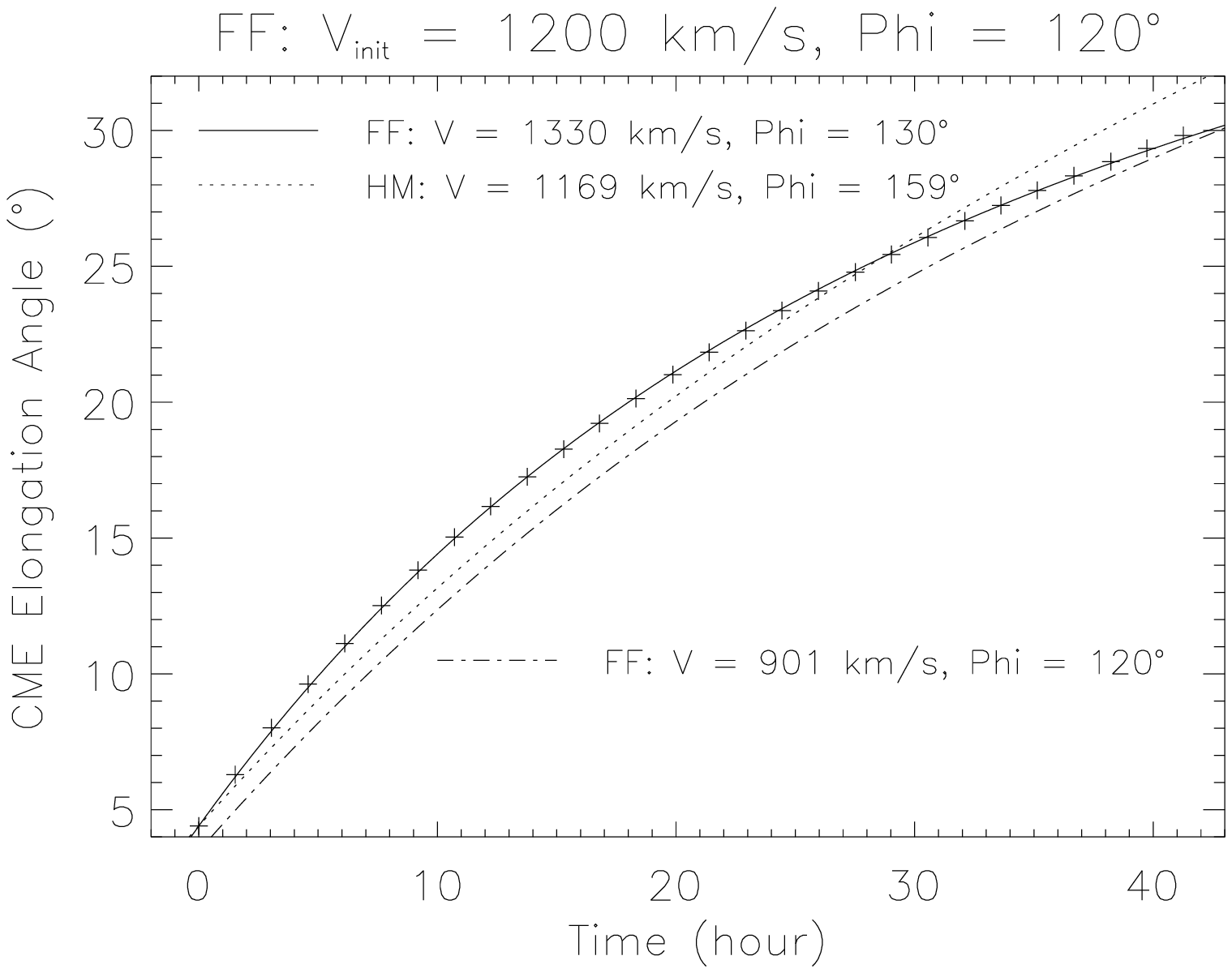}}
{\includegraphics*[width = 8cm]{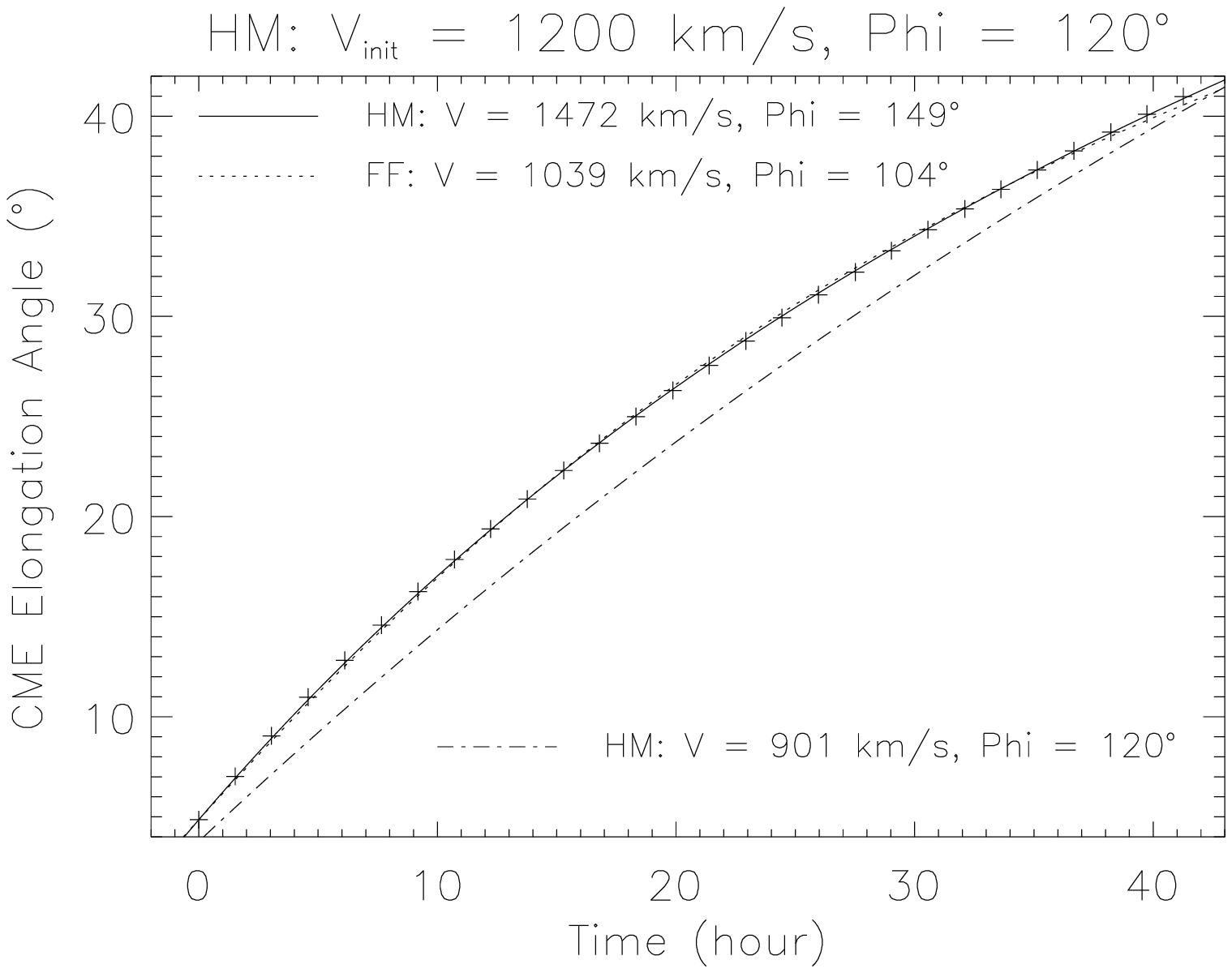}}
\caption{Top: Sample points (cross) corresponding to a time-elongation profile for an event where the elongation angle is given by the F$\Phi$ approximation, the distance is as shown in Figure~1 and the propagation angle is 120$^\circ$. The solid and dotted lines show the best-fit with the F$\Phi$ fitting and HM fitting methods, respectively and the dot-dashed line shows the profile created with the exact direction and average speed of the CME using the F$\Phi$ approximation. Bottom: Same as top but for a profile created with the HM approximation.  Here, the solid and dotted lines show the best-fit with the HM fitting and F$\Phi$ fitting methods, respectively and the dot-dashed line show the profile created with the exact direction and average speed of the CME using the HM approximation.}
\end{center}
\end{figure*}

Figure~2 shows the time-elongation points derived from time-distance profile using the F$\Phi$ (left) and HM (right) formulae. For the profile constructed with the F$\Phi$ method, the best-fit using the F$\Phi$ fitting method is 1330~$\pm$ 64~km~s$^{-1} $ and a direction of 130$^\circ \pm 1^\circ$ with respect to the Sun-spacecraft line. The best fit for the same profile but with the HM fitting method is 1169~$\pm$ 264~km~s$^{-1} $ and a direction of 159$^\circ \pm 14.5^\circ$ with respect to the Sun-spacecraft line.

For the profile constructed with the HM method, the best-fit using the HM fitting method is 1472~$\pm$ 90~km~s$^{-1}$ and a direction of 149$^\circ \pm 3^\circ$.  For the same profile, the best-fit with the F$\Phi$ fitting method is 1039~$\pm$ 104~km~s$^{-1}$ and a direction of 104$^\circ \pm 4.5^\circ$. In both panels, we also plot the ``expected'' profile for a CME with the same average speed and the same direction as the one used to create the profile. 

\begin{table}[htb]
\begin{tabular}{cccc}
\hline
CME Initial Speed &  Final Speed & Average Transit Speed & Transit Time (hours)\\
\hline
200 & 260 & 238 & 158.2\\
500 & 410 & 441 & 85.4\\
800 & 559 & 639 & 58.9 \\
1200 & 759 & 901 & 41.8 \\
1500 & 1058 & 1295 & 29.1 \\
\hline
\end{tabular}
\caption{Initial, final and average speed as well as transit time from 20~$R_\odot$ to 1~AU for the 5 synthetic CMEs in our study. All the speeds are in km~s$^{-1}$.}
\end{table}

It is clear from this case that CME deceleration has a considerable influence on the results of the two fitting procedures. For the fitting by the same method used to produce the tracks (solid lines), the results can be understood as follows: the physical deceleration due to interaction with the solar wind adds up to the geometrical deceleration due to the large propagation angle and yields too large directions.  Because the CME is ``seen'' as propagating farther away from the observing spacecraft than in reality, it also appears to propagate faster than in reality, especially to match the measurements closest to the Sun when the geometrical deceleration does not affect much the shape of the curve. While this reasoning holds true for both the F$\Phi$ and HM fitting methods, it has a stronger impact on the HM fitting. This is because for a same observed deceleration, the HM fitting returns a larger propagation angle as compared to the F$\Phi$ fitting \cite{Lugaz:2010c,Davies:2012}. Therefore, the physical deceleration (which is the same for the HM and F$\Phi$ cases) has a stronger influence on the results of the HM fitting procedures. 

It is also interesting to discuss the cases of ``cross-fitting'', i.e. fitting with one method a profile created using the other method (dash lines in Figure~2). In both cases, these fits give the best estimate of the average propagation speed of the CMEs. Specifically, for the profile created with the HM approximation, two of the effects described above balance each other: the added physical deceleration and the fact that, for a given angle, the HM fitting method expects less geometrical deceleration than the F$\Phi$ fitting method. It shows that, for fast and wide CMEs observed from large viewing angles, the F$\Phi$ fitting method of \inlinecite{Rouillard:2008} may work best because two types of errors associated with the fitting method cancel each other. 

We finish the analysis by discussing the expected arrival times with respect to the first observation at 20~R$_\odot$. For the HM fitting method, we use the correction of \inlinecite{Moestl:2011}. For the profile created with the F$\Phi$ method, the F$\Phi$ fit gives an arrival time of 28.3 hours, off by more than 12 hours. For the same profile, the HM fit gives an arrival time of 41.5 hours, almost perfect. However, the HM fit would predict a very glancing blow (39$^\circ$ away from the actual direction of propagation). For the profile created with the HM method, the HM fit gives an arrival of 29.2 hours, off by more than 12 hours, whereas the F$\Phi$ gives an arrival time of 36.2 hours and a direction (16$^\circ$ away from the actual direction of propagation) consistent with a hit. For both profiles, the cross-fittings provide the most accurate arrival times, particularly for the F$\Phi$ fitting of the HM profile which also gives a direction consistent with the actual CME direction.

\subsection{Complete Results}

We apply the same procedure as described above to the 40 synthetic tracks corresponding to different initial speeds and different directions. Table~1 lists the initial, final and average transit speeds for the 5 different initial speeds as well as the transit time from 20~R$_\odot$ to 1~AU. The results of the fitting procedures are shown in Figure~3, Tables~2 and 3. The results for the arrival times (including for the cross-fittings) are shown in Figure~4.

\begin{table}[htb]
\begin{tabular}{|c|cc|cc|cc|cc|cc|}
\hline
 & \multicolumn{10}{|c|}{Initial Speed}   \\
\hline
\multirow{2}{*}{Dir.} & \multicolumn{2}{|c|}{{\bf 200}} & \multicolumn{2}{|c|}{{\bf 500}} & \multicolumn{2}{|c|}{\bf 800} & \multicolumn{2}{|c|}{\bf 1200} & \multicolumn{2}{|c|}{\bf 1800} \\
& Speed & Dir. & Speed & Dir. & Speed & Dir. & Speed & Dir. & Speed & Dir.\\
\hline
{\bf 15} & 234 & 13 & 445 & 16 & 651 & 18 & 925 & 19 & 1333 & 19\\
{\bf 30} & 232 & 26 & 450 & 33 & 667 & 36 & 953 & 37 & 1382 & 38\\
{\bf 45} & 228 & 40 & 458 & 49 & 692 & 53 & 1001 & 55 & 1470 & 57\\
{\bf 60} & 222 & 53 & 469 & 65 & 720 & 69 & 1063 & 72 & 1578 & 74\\
{\bf 75} & 215 & 67 & 484 & 81 & 758 & 85 & 1136 & 88 & 1704 & 90\\
{\bf 90} & 206 & 81 & 497 & 96 & 795 & 100 & 1213 & 103 & 1801 & 104\\
{\bf 105} & 198 & 96 & 515 & 111 & 846 & 115 & 1284 & 117 & 1923 & 118\\
{\bf 120} & 191 & 112 & 524 & 125 & 861 & 128 & 1331 & 130 & 2013 & 131\\
\hline
\end{tabular}
\caption{\underline{F$\Phi$ Fitting Results}: \newline Best-fit speed and direction for the 40 synthetic tracks constructed and fitted with the F$\Phi$ approximation. All the speeds are in km~s$^{-1}$ and all the angle in $^\circ$ with respect to the Sun-spacecraft line.}
\end{table}

First, we discuss the error in the best-fit direction (top panel of Figure~3). For both fitting methods, for CMEs faster than the solar wind speed, the error in the direction increases with increasing speed. This is expected as faster CMEs experience more drag and there is a larger physical deceleration added to the geometrical deceleration. For the case when the CME initial speed is 200~km~s$^{-1}$, the errors in direction are negative (actual direction is underestimated by the fitting methods). As is clear from this Figure and Tables~2 and 3, the error in the direction of propagation is typically around 10$^\circ$--15$^\circ$ for the F$\Phi$ method and 20$^\circ$--30$^\circ$ for the HM method.  

\begin{figure*}[t*]
\begin{center}
{\includegraphics*[width = 9cm]{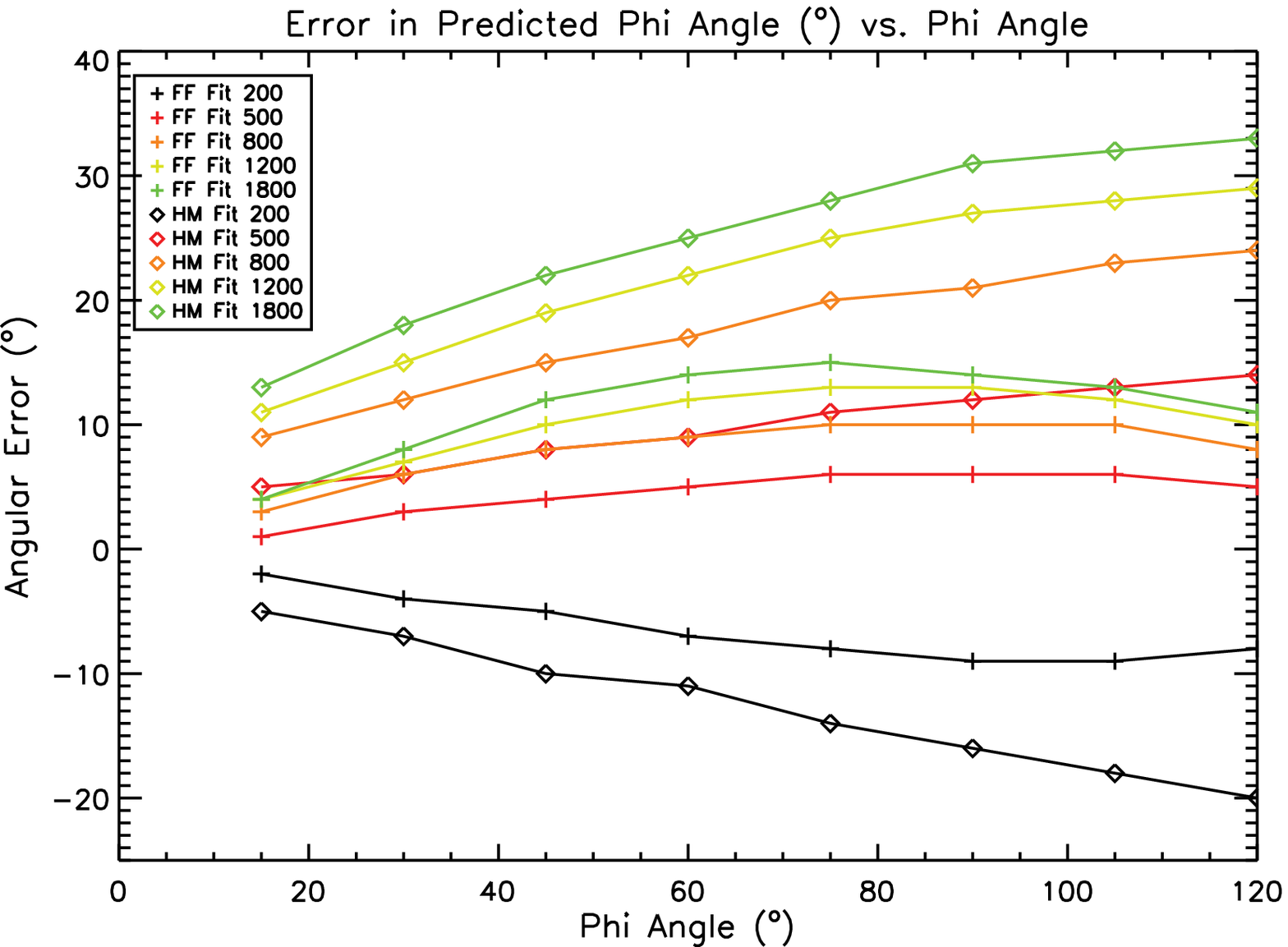}}
{\includegraphics*[width = 9cm]{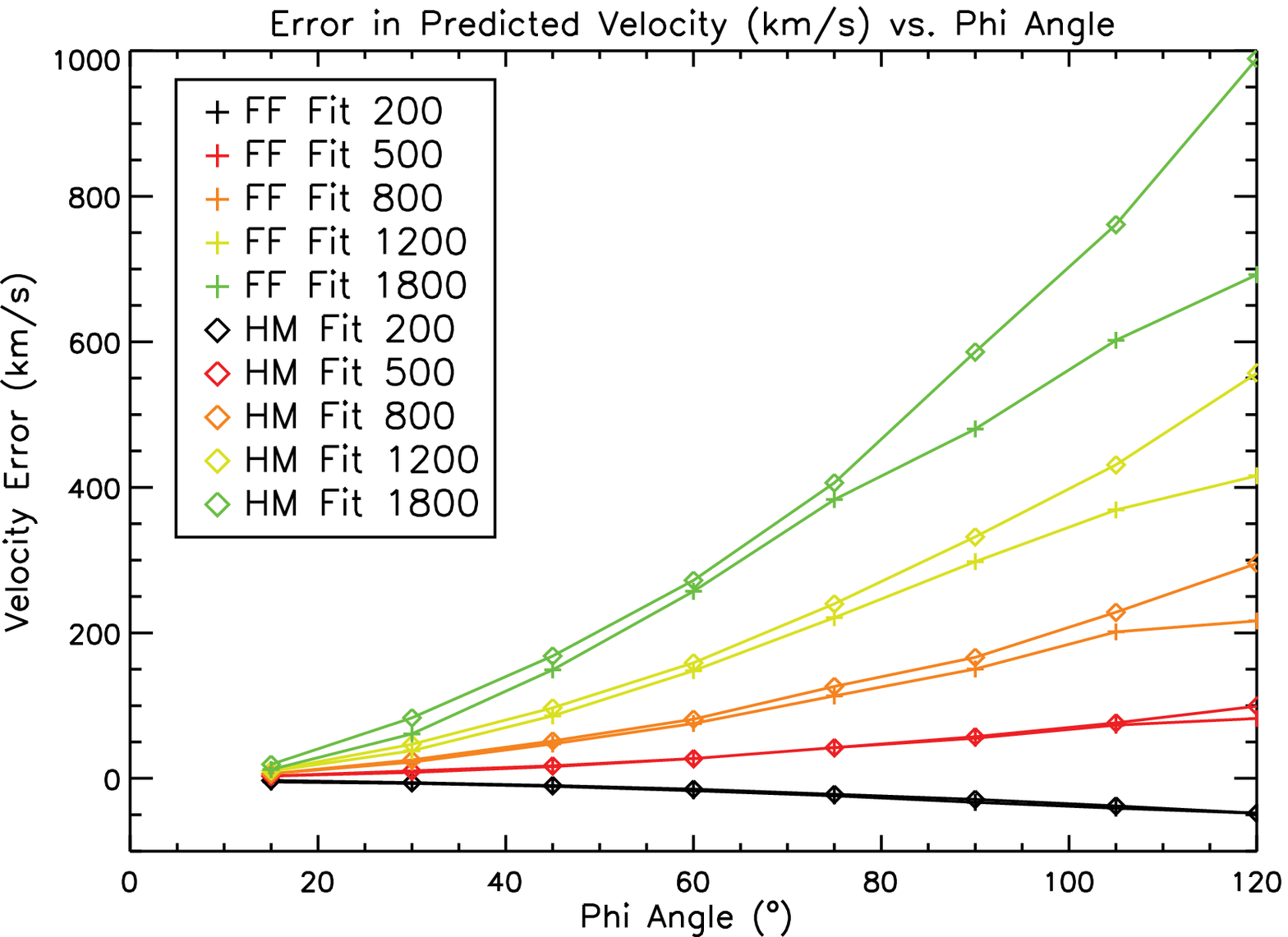}}
\caption{Top: Difference between the best-fit direction and the input direction as a function of the input propagation direction. Colors refer to different input speeds; crosses and diamonds are used for profiles created with the F$\Phi$ and HM approximation, respectively. Bottom: Same as top but for the difference between the average transit speed and the best fit speed (as listed in Tables 1, 2 and 3). Negative values correspond to cases when the fitted results underestimate the actual parameters.}
\end{center}
\end{figure*}

Regarding the error in velocity (bottom panel of Figure~3), we plot the error with respect to the average transit speed of the CME. For both fitting methods, the error in the average velocity increases with increasing viewing angles and, also, with larger difference with the solar wind speed. Faster CMEs experience more drag and are seen as propagating farther away from the observing spacecraft and therefore, even faster than in reality. Also, larger viewing angles correspond to stronger geometrical deceleration yielding larger errors. Overall, as could be expected, the best-fit velocity is closer to the average transit speed of the CME than it is to the final speed. Using the best-fit speed without correction to predict the measured (final) speed of the CME can only work for CMEs with initial speed close to that of the solar wind. For CMEs which experience deceleration/acceleration due to their interaction with the solar wind, it is necessary to apply corrections or to assume a varying speed (see also \opencite{Rollett:2011}). Here, we find that slow CMEs with initial speed of 200~km~s$^{-1}$ and 500~km~s$^{-1}$ observed front-sided (actual direction of 90$^\circ$ or less) have a best-fit velocity between the initial and final speed of the CME, and this for both fitting methods. For this type of CMEs, which correspond to the vast majority of events observed so far by STEREO, the two fitting methods can be used to predict the CME final speed with reasonable error. This is not the case for faster CMEs observed behind the limb, for which the best-fit velocity may be as large as twice the final speed of the CME. 

\begin{table}[tb]
\begin{tabular}{|c|cc|cc|cc|cc|cc|}
\hline
 & \multicolumn{10}{|c|}{Initial Speed}   \\
\hline
\multirow{2}{*}{Dir.} & \multicolumn{2}{|c|}{{\bf 200}} & \multicolumn{2}{|c|}{{\bf 500}} & \multicolumn{2}{|c|}{\bf 800} & \multicolumn{2}{|c|}{\bf 1200} & \multicolumn{2}{|c|}{\bf 1800} \\
& Speed & Dir. & Speed & Dir. & Speed & Dir. & Speed & Dir. & Speed & Dir.\\
\hline
{\bf 15} & 236 & 10 & 445 & 20 & 651 & 24 & 926 & 26 & 1340 & 28 \\
{\bf 30} & 233 & 23 & 452 & 36 & 670 & 42 & 962 & 45 & 1404 & 48 \\
{\bf 45} & 229 & 35 & 459 & 53 & 696 & 60 & 1012 & 64 & {\it 1489} & {\it 67} \\
{\bf 60} & 224 & 49 & 469 & 69 & 726 & 77 & {\it 1074} & {\it 82} & {\it 1593} & {\it 85} \\
{\bf 75} & 217 & 61 & 484 & 86 & 771 & 95 & {\it 1155} & {\it 100} & {\it 1727} & {\it 103} \\
{\bf 90} & 210 & 74 & 499 & 102 & {\it 811} & {\it 111} & {\it 1247} & {\it 117} & {\it 1907} & {\it 121} \\
{\bf 105} & 201 & 87 & 518 & 118 & {\it 873} & {\it 128} & {\it 1346} & {\it 133} & {\it 2082} & {\it 137} \\ 
{\bf 120} & 191 & 100 & 541 & 134 & {\it 940} & {\it 144} & {\it 1472} & {\it 149} & {\it 2310} & {\it 153} \\
\hline
\end{tabular}
\caption{\underline{HM Fitting Results}: \newline Best-fit speed and direction for the 40 synthetic tracks constructed and fitted with the HM approximation. Results in italic are those for which the error in direction is greater than 20$^\circ$. All the speeds are in km~s$^{-1}$ and all the angle in $^\circ$ with respect to the Sun-spacecraft line.}
\end{table}

\subsection{Consequences for Predicting CME Hit and Arrival Time}

We now discuss the consequences of the errors in the direction and velocity to predict the CME hit/miss at Earth (or at one of the STEREO spacecraft) and to predict the arrival time (or equivalently the total transit time). The F$\Phi$ method assumes a negligible CME width \cite{Rouillard:2008}. On the opposite, the HM method assume a very wide width, so that a CME propagating  90$^\circ$ away from the Sun-Earth line is expected to hit Earth. In previous studies \cite{Davis:2011,Lugaz:2012a}, CMEs with direction given by any of the two methods within $\pm 15^\circ$--20$^\circ$ of the Sun-Earth line have been considered as likely to hit Earth. This takes into consideration the non-zero width of the CMEs and the typical uncertainties in the best-fit direction. Hereafter, we use $\pm 20^\circ$ error as the limit for a correct prediction of the hit/miss of a CME from the two fitting methods.

Regardless of the speed and the direction of the CME, the error in the direction for tracks created using the Fixed-$\Phi$ approximation is less than 15$^\circ$. This implies that the effect of solar wind drag would be relatively negligible to make accurate prediction of the CME hit/miss if the Fixed-$\Phi$ approximation is the correct one to describe a CME. On the contrary, for a fast and wide CME (with initial speed greater than 800 km~s$^{-1}$) propagating towards Earth more than 90$^\circ$ from the observing spacecraft, a miss would be mistakenly predicted by the HM fitting method. A fast and wide CME propagating more than 20$^\circ$ east of the Sun-Earth line would be predicted by the HM fitting method to hit Earth. These cases when the HM method would fail in its prediction of hit/miss are shown in italic in Table~3.

We then turn our focus on the predicted arrival time. Following previous works, we consider that, under the F$\Phi$ approximation, each part of the CMEs within $\pm 20^\circ$ of the best-fit direction have the best-fit speed, whereas, under the HM approximation, we use the corrected speed given by \inlinecite{Moestl:2011}. We derive the predicted arrival time even for CMEs with best-fit directions more than 20$^\circ$ away from the actual direction. For CMEs faster than the solar wind speed, the HM fits result in lower errors in the arrival times than the F$\Phi$ fits, even though the direction is often off by more than 20$^\circ$. The typical error in the transit time associated with neglecting the change in speed caused by the solar wind drag is about 5--15 hours. Since the errors are relatively constant independently of the initial speed of the CME, the relative error is large for the fastest CME (up to 30$\%$). For the slowest CME (200~km~s$^{-1}$ initial speed), the relatively small errors in the velocity correspond to large errors (more than 2 days) for the predicted arrival time, also corresponding to about 30$\%$ relative error. While errors of $\sim$ 12 hours are typical for models designed to predict CME arrival times \cite{Davis:2011}, it should be understood that there are additional source of errors not captured here, such as the fact that no CME can be exactly described by the HM or F$\Phi$ approximation, which could make the typical error in the arrival time even larger. 

\begin{figure*}[t*]
\begin{center}
{\includegraphics*[width = 9cm]{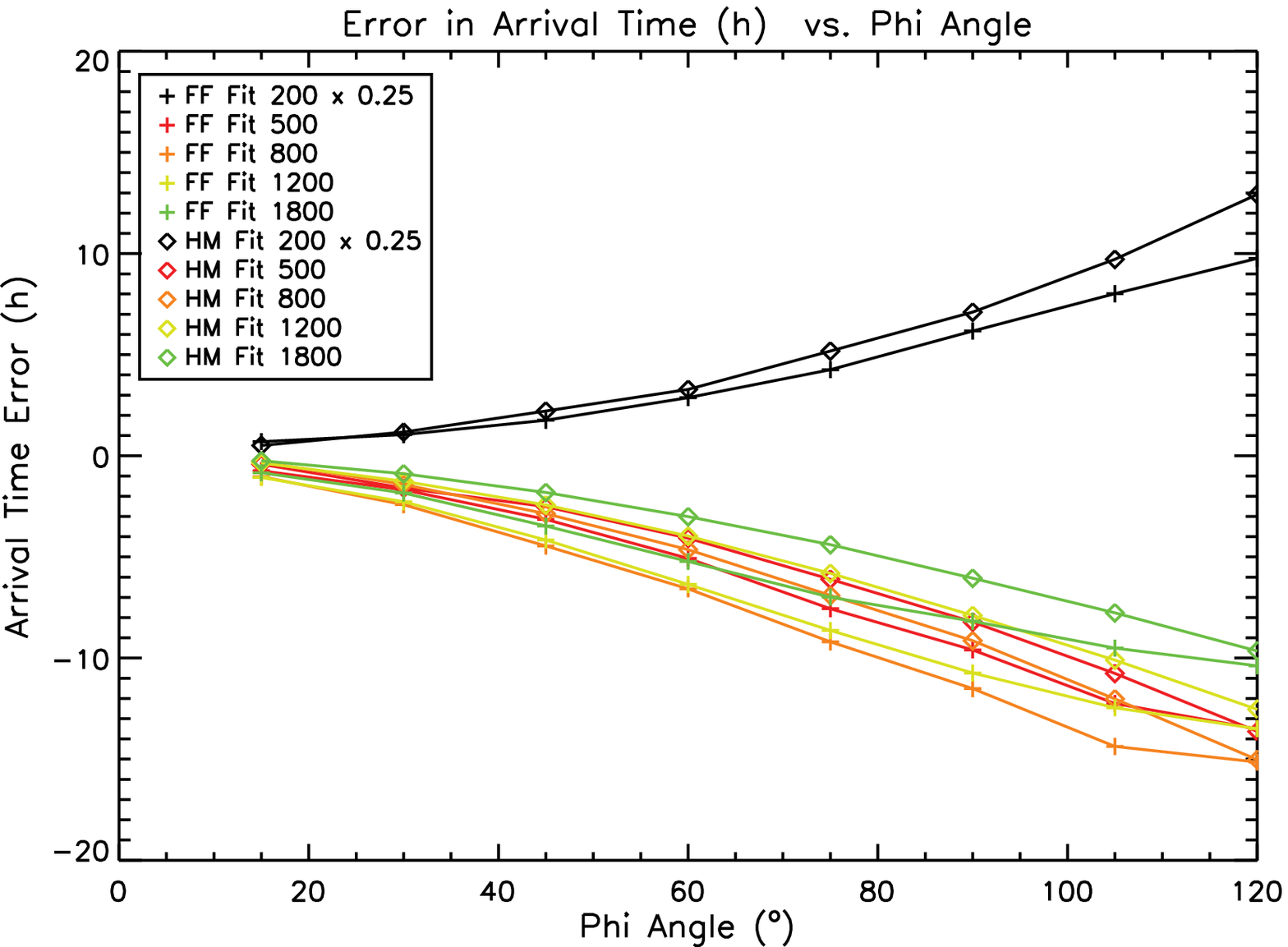}}
{\includegraphics*[width = 9cm]{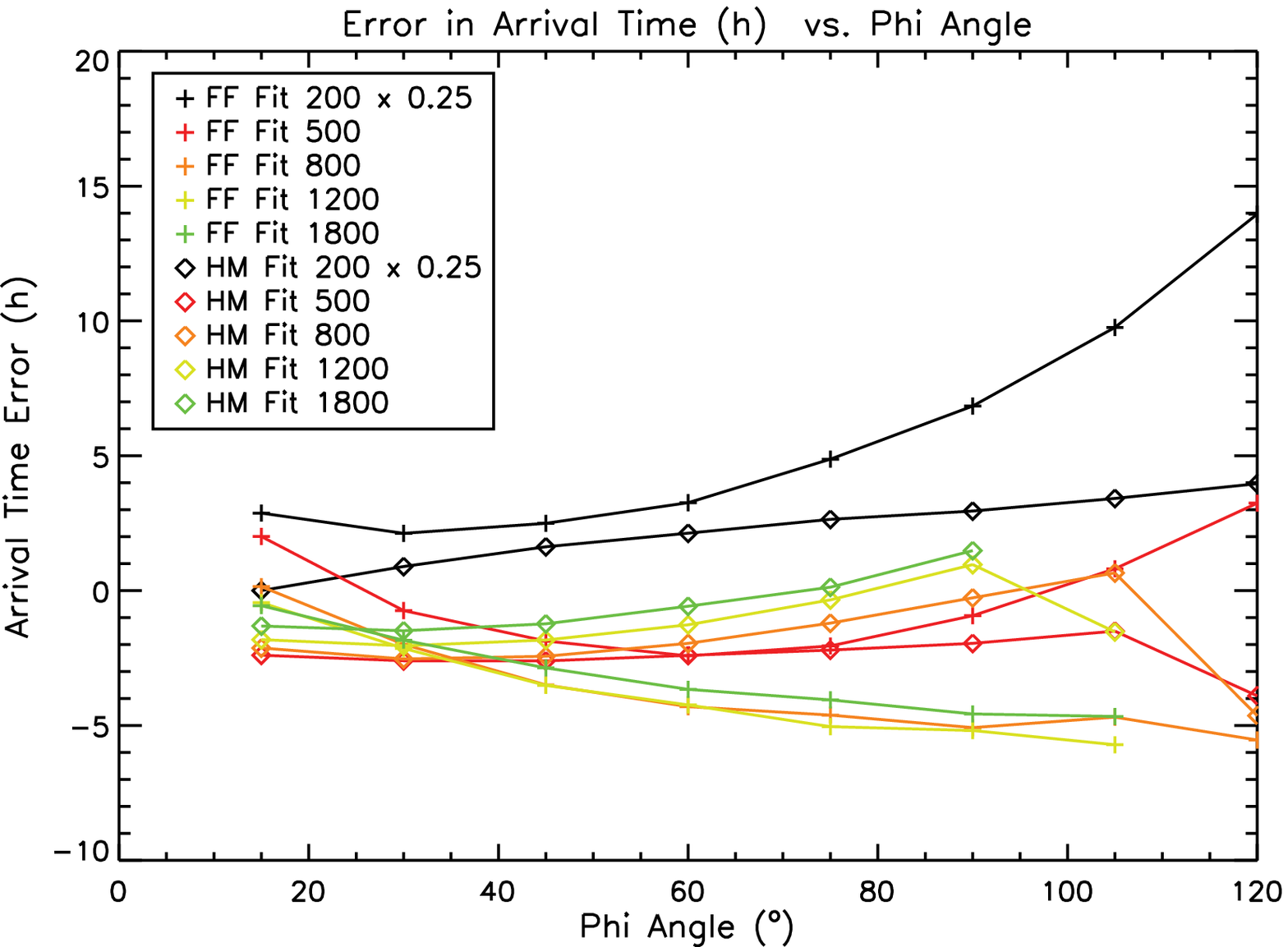}}
\caption{Top: Difference between the best-fit transit time and the actual transit time as a function of the input propagation direction. Colors refer to different input speeds; crosses and diamonds are used for profiles created with the F$\Phi$ and HM approximation, respectively. Bottom: Same as top but for cross-fitting, the legend refers to the method used to fit the profiles. In both panels, the results for the profile with initial speed 200~km~s$^{-1}$ are divided by four.}
\end{center}
\end{figure*}

Finally, we quickly discuss the errors for the cross-fitting (bottom panel of Figure~4) since the F$\Phi$ fit of the HM profile may yield the closest match to the CME direction and average speed as was discussed in section 3.1. We find that, in general, these fits result in errors in the arrival times about a factor of 2 lower than that from the direct fits. In effect, two phenomena cancel each other: the HM profiles have less deceleration than F$\Phi$ profiles and the physical deceleration adds up to the geometrical deceleration. Because of this, synthetic profiles created using the F$\Phi$ approximation are best-fitted with the HM fitting method and reciprocally. This result even holds true for the CMEs which experience the least interaction with the solar wind (200 and 500~km~s$^{-1}$). It implies that, for real CME tracks, it might be impossible to distinguish between the effect of the non-zero width and the deceleration as these two effects tend to cancel each other. Previous studies have shown that the F$\Phi$ and HM methods give similar best-fit values except for CMEs propagating towards the observing spacecraft or behind the limb \cite{Lugaz:2010c,Davies:2012}. Here, we find that even for CMEs propagating behind the limb, the effect of the non-constant CME speed is to make it nearly impossible to assess which model bests represents the CME actual geometry.

\section{Discussions and Conclusions}

In this paper, we have studied the influence of the interaction between coronal mass ejections (CMEs) and the solar wind to determine the properties (direction, speed, transit time) of CMEs reconstructed by applying fitting methods of HI observations. As CMEs propagate through the inner heliosphere, their speed gets closer to that of the background solar wind and it should result in errors in any method, which assumes a constant CME speed. The two main methods of fitting time-elongation profiles obtained from a single HI instruments are the F$\Phi$ and HM fitting methods, both of these assume a constant velocity. Here, a simplified model for CME propagation through the inner heliosphere has been used to quantify the error that real time-elongation profiles may have due to the interplanetary aerodynamic drag. We have found that the best-fit speed cannot be used in general to represent the average transit speed of the CME or its final speed at 1~AU as it is typically done for slow CMEs. This is particularly true for fast CMEs observed behind the limb, which is the expected STEREO-viewing geometry for geo-effective events in 2012. 

In addition, we have found that the physical deceleration of fast CMEs can result in errors in the estimated direction of propagation as large as 15$^\circ$ and 30$^\circ$ for the F$\Phi$ and HM methods respectively. Overall the F$\Phi$ fitting method does a better job at predicting the direction of propagation and the average velocity than the HM method when the CME varying speed is  included. This difference is barely noticeable when comparing predicted velocities at small viewing angles and slow initial velocities, but is much greater at higher viewing angles and  faster initial velocities. However, the error in the arrival time (or transit time) is nearly identical for the two methods and can reach as high as 12 hours for average to fast CMEs. This is comparable to the errors for existing models \cite{Davis:2011} and imply that existing methods may need to be further refined in order to perform real-time space weather forecasting for the fastest CMEs. We also found that the error associated with mis-characterizing the CME geometry (for example neglecting its width or assuming a too large width) typically cancels the error associated with neglecting the CME heliospheric deceleration. In previous works \cite{Lugaz:2010c,Lugaz:2012a,Davies:2012}, we have tried to compare the F$\Phi$ and HM methods to determine for which viewing conditions and types of CMEs one approximation is more valid than the other. The present work tends to question whether such a conclusion can be reached, since we have found that a wide, decelerating CME may be best fitted using an approximation assuming a negligible CME width.

It should be noted that the model used for the solar wind drag, as described in Equation 4, is not the most sophisticated. It coalesces all effects associated with the CME surface area, the CME mass and the solar wind density into a radial dependence of $R^{-\frac{1}{2}}$. In addition, the solar wind speed was assumed to be constant equal to 320~km~s$^{-1}$ between 0.1 and 1~AU. We believe these two assumptions are still satisfactory for creating time-elongation tracks, and that the main conclusions of our studies would hold with different models of the interaction between the solar wind and CMEs (non-constant speed, snow-plow, etc...). Assuming a constant speed is not the only contributor to the error for fitting methods but is part of it. Here, we have been able to quantify this error to show how it contributes to the prediction of the arrival at 1~AU of CMEs. A better fitting method might include a velocity that changes according to physical models for CME kinematics, although more work is required to determine the best model for solar wind drag or snow-plow and the associated coefficients following studies such as those by \inlinecite{THoward:2007}, \inlinecite{Webb:2009}, \inlinecite{Byrne:2010}, \inlinecite{Maloney:2010}, \inlinecite{Vrsnak:2010} and \inlinecite{Temmer:2011}.

 \begin{acks}
N.~L. was supported during this work by NSF grants AGS 0819653, 0639335 and 1140411 and NASA grants NNX08AQ16G and NNX12AB28G. P.~K. performed research for this work under a Research Experience for Undergraduates (REU) position at the University of Hawaii's Institute for Astronomy and funded by NSF. 
SOHO and STEREO are projects of international cooperation between ESA and NASA. 
   The SECCHI data are produced by an international consortium of 
  Naval Research Laboratory, Lockheed
  Martin Solar and Astrophysics Lab, and NASA Goddard Space Flight
  Center (USA), Rutherford Appleton Laboratory, and University of
  Birmingham (UK), Max-Planck-Institut f{\"u}r Sonnensystemforschung
  (Germany), Centre Spatiale de Liege (Belgium), Institut d'Optique
  Th{\'e}orique et Appliqu{\'e}e, and Institut d'Astrophysique
  Spatiale (France).  
\end{acks}

\bibliographystyle{spr-mp-sola-cnd}

\end{article}

\end{document}